\documentstyle[aps,twocolumn]{revtex}
\input{psfig}
\begin{document}
\draft
\title{Metastable Voltage States of Coupled Josephson Junctions}
\author{Ya.~M.~Blanter$^{a,b}$, Gerd Sch\"on$^{c}$, and
A.~D.~Zaikin$^{c,d}$} 
\address{$^a$ Institut f\"ur Theorie der Kondensierten Materie,
Universit\"at Karlsruhe, 76128 Karlsruhe, Germany\\
$^b$ Department of Theoretical Physics, Moscow Institute for Steel and
Alloys, Leninskii pr. 4, 117936 Moscow, Russia\\
$^c$ Institut f\"ur Theoretische Festk\"orperphysik,
Universit\"at Karlsruhe, 76128 Karlsruhe, Germany\\
$^d$ I.~E.~Tamm Department of Theoretical Physics, P.~N.~Lebedev
Physics Institute, Leninskii Pr. 53, 117924 Moscow, Russia}
\date{\today}
\maketitle 
\tighten

\begin{abstract}
We investigate a chain of capacitively coupled
Josephson junctions in the regime where the charging
energy dominates over the Josephson coupling, exploiting the analogy
between this system and a multi-dimensional crystal. We find that
the current-voltage characteristic of the current-driven chain 
has a staircase shape, beginning
with an (insulating) non-zero voltage plateau at small currents.
This behavior differs qualitatively from that of a single junction,
which should show Bloch oscillations with vanishing dc voltage. 
The simplest system where this effect can be observed
consists of three grains connected by two junctions. The theory
explains the results of recent experiments on Josephson junction
arrays.  
\end{abstract}
\pacs{PACS numbers: 73.23.-b,74.50.+r}

Recent experiments \cite{Jap} carried out on two-dimensional Josephson
junction arrays in the quantum regime -- 
i.e. where the charging energy exceeding the
Josephson coupling introduces quantum dynamics -- 
showed remarkable steps of the current-voltage characteristics. 
Two voltage steps could be clearly seen,
with voltage value given by the superconducting gap  $2\Delta/e$.
 Further structure at higher voltages is washed out. 
The purpose of the present article is to show that these steps can be 
naturally explained in the framework of the Bloch oscillation description
of quantum mechanical Josephson junctions. 
Moreover, we argue that the physics of a chain differs qualitatively 
from that known from single Josephson junctions  \cite{Likh,SZ}. 
In contrast to a  current-biased single-junction, a chain of 1D Josephson
junctions has no
zero-voltage state, i.e. the average voltage is greater than $2\Delta/e$
for any current. The only stable state of the system has a voltage equal to
$2N\Delta/e$ where $N$ is the number of junctions. However, for low
currents this stable state is reached only on astronomical time
scales,
$$t_Z = 2eI^{-1} \exp(I_Z/I),$$
while the time required for the voltage $2\Delta/e$ to establish is 
$$t_s \sim 2e\Delta/IE_C \ll t_Z.$$
Here $E_C$ is the charging energy, and $I_Z$ is defined below. 
Thus, in realistic experiments metastable 
$I$--$V$ characteristic are observed, with voltage steps starting from the 
value $2\Delta/e$. The time required to achieve the stable 
 voltage depends strongly on the  external current. 
These results are a consequence of a lowered
symmetry of the system, and the effect shows up already in the system
shown on Fig.~1. We conjecture that the
physics of 2D Josephson junction arrays is similar, which provides an
explanation of the experiments \cite{Jap}.  

First we recollect the results of a single Josephson junction with a
capacitance $C$ and Josephson coupling energy $E_J$. In the
absence of dissipation and for energies below the superconducting
energy gap $\Delta$ it can be described by the Hamiltonian \cite{Likh,SZ}
\begin{equation} \label{ham}
\hat H = \frac{\hat Q^2}{2C} - E_J \cos \phi, \ \ \ \hat Q = -2ie
\frac{\partial}{\partial \phi},
\end{equation}
where $\hat Q$ and $\hat \phi$ are the charge and the 
superconducting phase difference operators,  respectively. 
The Hamiltonian (\ref{ham})
is equivalent to that of a quantum particle with a coordinate 
$\phi$ in a periodic potential. If the junction is driven 
by an external current $I$
(which is an analogue of an external force for a quantum particle),
the junction charge  increases with time, $Q=It$, 
until it reaches the value $Q=e$. Then the system has two options: 
either a Cooper pair tunnels across the junction
and the junction charge is reset from $Q=e$ to $Q=-e$, or if this does not
happen the charge increases further with time. The former option 
is analogue to a Bragg reflection when a quantum particle
reaches the edge of the Brillouin zone and is reflected, thus staying 
in the lowest energy band. The latter option is the equivalent of Zener 
tunneling across the band gap 
to a higher band \cite{Ziman}. In the limit $E_J \ll E_C = e^2/2C$ 
(which will be discussed in this paper) the probability of
such Zener tunneling event is \cite{Ziman}
\begin{equation} \label{pro1}
\lambda = \exp(-I_Z/I),\;\;\; I_Z=\pi e E_J^2/8E_C.
\end{equation}
The question is which of the two options are realized and how they
manifest themselves in an experiment. The answer depends on the current:

{\it Low currents}: In this regime the average time $t_Z$ required 
for a Zener tunneling event to occur exceeds the time of the experiment,
 $t_Z = 2e(I\lambda)^{-1} \gg t_e$. This means that during the whole
experiment the system stays in the lowest energy band, the charge 
(and the voltage) across the junction oscillates with the period $2e/I$ 
around zero. This is the metastable regime of Bloch oscillations
\cite{Likh,SZ,Lyonya}. 

{\it High currents}: In this regime, $t_e \gg t_Z$, Zener
tunneling event \cite{Zener} takes place during the experiment. In the
absence of dissipation the system  -- after it jumps to the second 
energy band --
will continue to rise in energy \cite{foot1} until its energy reaches
the value $2\Delta +E_C$ (the average voltage across the junction at
this point is $V_0 \approx 2\Delta /e$ \cite{SZ}). A further increase
is not possible because of dissipation due to single electron
tunneling that sets in as soon as the voltage exceeds the value 
$2\Delta /e$. Then the
voltage will oscillate around the stable value $2\Delta /e$ \cite{SZ}.

Thus the difference between the two above regimes is quite clear:
the first one corresponds to metastable zero-voltage state of the
junction whereas in the second regime a finite stable voltage $V
\approx 2\Delta /e$ is  measured. The crossover between these two
regimes occurs at $I \approx I_Z / \ln (It_e/2e)$.

Experimentally it is more convenient to study charging effects
in systems containing several (or many) Josephson junctions. 
Hence one can ask
whether the simple physical picture discussed above remains valid also
in the case of two or more coupled Josephson junctions. Intuitively
one could expect that for  sufficiently low currents flowing through the
chain of Josephson junctions the system will stay
in the zero voltage state $\bar V \approx 0$ similarly to the case of
a single junction.

In this paper we argue that the behavior of already two capacitively 
coupled Josephson junctions  is {\em qualitatively different}: 
a finite voltage drop $V \simeq V_0 \equiv 2\Delta /e$ across the
system will be  
measured {\it no matter how low the external current is}. This 
conclusion remains
valid for an arbitrary number of junctions in the chain $N>1$. To
understand this result we use the analogy between
our system and an electron in a multidimensional crystal. If
the self-capacitance $C_0$ of the superconducting islands
is non-zero the symmetry of the corresponding Brillouin
zone is lowered in a way which prevents the system from staying in the 
lowest energy band for {\it any} non-zero current $I$. Zener tunneling in
at least one of the junctions occurs immediately after the current is
applied and the system switches to a finite voltage state.
The resulting current-voltage characteristics of the Josephson
junction chain has a form of a staircase. Our results 
can explain the low voltage behavior of two-dimensional Josephson junction
arrays recently reported in Ref. \cite{Jap}.

\begin{figure}
\centerline{\psfig{figure=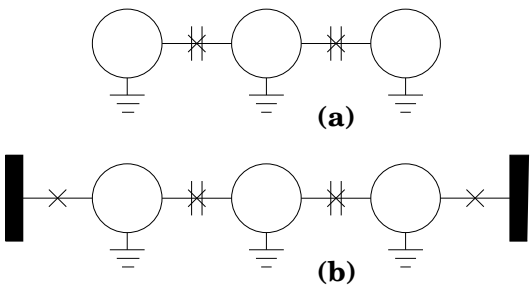,width=7.cm}}
\vspace{0.3cm}
\caption{(a) Two capacitively coupled Josephson junctions; (b) An
experimental realization of the same system.} 
\label{picture}
\end{figure}

For the sake of simplicity we first consider the system of three 
superconducting grains with
phases $\phi_i$, $i=1,2,3$, connected by two Josephson junctions
(Fig.~1a). The 
relevant variables for description of this system are linear
combinations of the phases (cf. e.g. \cite{Brink})
$$\theta_1 = \phi_1-\phi_2; \ \ \ \theta_2 = \phi_2 - \phi_3; \ \ \
\theta_3 = (\phi_1 + \phi_2 + \phi_3)/3.$$ 
We can also introduce the junction charge operators
$\hat Q_i = -2ie\partial/\partial \theta_i$, $i=1,2$, and the
``center-of-mass charge'' $\hat Q_3 = -2ie\partial/\partial
\theta_3$. Then the Hamiltonian reads as
\begin{equation} \label{ham1} 
\hat H =  \frac{\hat Q_3^2}{6C_0} + \frac{1}{2}
\sum_{ij} \hat Q_i (\hat C^{-1})_{ij} \hat Q_j - E_J \sum_i \cos
\theta_i,  
\end{equation}
and the effective 2$\times$2 capacitance matrix $\hat C$ has elements
$C_{11} = C_{22} = C + 2C_0/3$; $C_{12} = C_{21} = C_0/3$.
The center-of-mass dynamics is independent of 
that of both junctions. The latter is essentially 
the motion of an electron in a 2D periodic
potential. In the limiting case $E_J \ll e^2/\max\{C_0,C\}$
(nearly-free-electron model) the Josephson term in the Hamiltonian
(\ref{ham1}) can be considered as a perturbation. We assume furthermore
that $2\Delta > e^2/\max\{C_0,C\}$. The eigenfunctions of the unperturbed
Hamiltonian are ``plain waves'',
$$\psi_0(\mbox{\boldmath $\theta$ \unboldmath}) =
\exp(i\bbox{Q}\mbox{\boldmath $\theta$ \unboldmath}/2e +
iQ_3\theta_3/2e).$$  
Here we have introduced two-dimensional vectors $\mbox{\boldmath
$\theta$ \unboldmath} = (\theta_1,\theta_2)$ and $\bbox{Q} =
(Q_1,Q_2)$.  

Now we can clarify the meaning of the charges $Q_1$ and
$Q_2$. Introducing the grain charge operators $\hat q_i =
-2ie\partial/\partial \phi_i$, $i=1,2,3$, with the eigenvalues $q_i$,
we obtain
$$q_1 = Q_1 + Q_3/3; \ q_2 = -Q_1 + Q_2 + Q_3/3; \ q_3 = -Q_2
+ Q_3/3.$$
The charge $Q_3$ is just a total charge of the
chain. It is conserved, and from now on we put it equal to zero
\cite{foot2}. In the current-bias regime the charge
$q_1$ grows linearly with time, except for the jumps by $2e$ due to
Cooper pair tunneling (CPT) through junction 1 (see below). The
charge $q_3$ decreases linearly with time, except for
CPT through junction 2, while the charge $q_2$ keeps constant
except for CPT through the either junction. In other words, the system
can move only along lines $Q_1 - Q_2 = 2en$, with $n$ being an
integer, and CPT processes are responsible for the jumps of the system
between these lines. The time evolution of the system starts from the
line $Q_1 = Q_2 = It$.  

The eigenvalues of the unperturbed Hamiltonian are 
\begin{equation} \label{pov}
E_0(\mbox{\bf Q}) =  \frac{1}{2} \sum_{ij} Q_i (\hat C^{-1})_{ij} Q_j.
\end{equation} 
The Josephson coupling plays a role only in the vicinity of ``critical
surfaces'', $E_0(\mbox{\bf Q}) = E_0(\mbox{\bf Q} - \mbox{\bf K})$, 
with $\mbox{\bf K} = (2pe,2eq)$ ($p$ and $q$ are integer numbers)
being the reciprocal lattice vectors. In lowest (second) order in
$E_J$ only vectors $\mbox{\bf K}$ with $p=\pm 1,q=0$ and $p=0, q=\pm
1$ are essential, and the Josephson coupling creates gaps equal to $E_J$
on the critical surfaces. For the spectrum (\ref{pov}) these are given
by 
$$\mbox{1)}\  Q_1 - \alpha Q_2 = \pm e; \ \ \ \mbox{2)} \ Q_2 - \alpha
Q_1 = \pm e,$$ 
with $\alpha = (C_0/3)(C + 2C_0/3)^{-1}$. Note that $0 \leq \alpha \leq
1/2$; the limits $\alpha = 0$ and $\alpha = 1/2$ correspond to
the cases $C_0=0$ and $C=0$, respectively. Thus, the critical
surfaces are just two pairs of parallel straight lines. The romb
formed by these lines is the first Brillouin zone. The Bragg
reflection processes are possible on the critical surfaces, i.e. the
system can jump by a vector $\mbox{\bf K}$. The energy is
periodic in $\mbox{\bf Q}$-plane with two orthogonal periods equal to
$2e$. It will be convenient for us to use the extended
band picture. The voltage across the junction $i$ is
given by $V_i = \partial E/\partial Q_i$.

\begin{figure}
\centerline{\psfig{figure=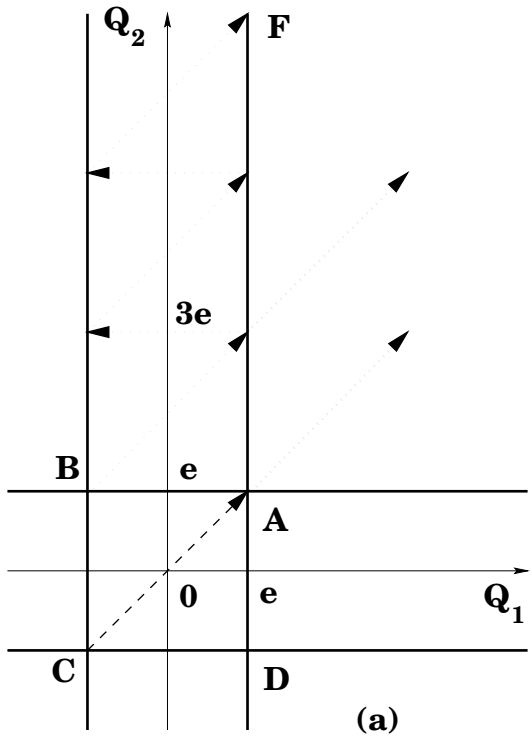,width=7.cm}}
\vspace{0.3cm}
\caption{The motion of the current-biased system in the plane
$(Q_1,Q_2)$; $C_0 = 0$. The critical surfaces are shown by boldface
lines; the first Brillouin zone is a square ABCD. In the weak current
regime the system moves along the line AC. Possible trajectories
for the cases of strong currents, $\lambda \sim 1$, are shown by
dotted lines.} 
\label{brill1}
\end{figure}

For $C_0=0$ the Hamiltonian (\ref{ham1}) is just a sum of two
Hamiltonians (\ref{ham}) of isolated junctions, and the behavior
of the chain is trivial. 
The critical surfaces are $Q_i = \pm e$, $i=1,2$ (Fig.~2). For low
currents a possible time evolution (with large probability
$(1-\lambda)^2$) is simultaneous  Bloch oscillations in two junctions
\cite{Geig}, i.e. Bragg reflection from point A$=(e,e)$ to point
C$=(-e,-e)$. The mean voltage is zero for this process. Another 
possibility (with probability $\lambda (1 - \lambda)$) is 
the reflection to the point B$=(-e,e)$, with 
subsequent motion along the line $(Q,Q+2e)$, $Q > -e$. This means
Zener tunneling in junction 2, and Bloch oscillations in 
junction 1. Eventually the system reaches the critical
surface in the point $(e,3e)$, where it is reflected to the point
$(-e,3e)$ and 
so on. The voltage across junction 1 $V_1 = Q_1/C$ oscillates 
with frequency $I/2e$, while the voltage across junction 2 grows
linearly in time. The third possibility (with equal probability
$\lambda(1-\lambda)$) is an equivalent process, where the
system is initially reflected to the point $(e,-e)$.
Finally, the last possibility (with probability $\lambda^2$) is
to proceed along the line $(Q,Q)$ without any
reflection. This corresponds to Zener tunneling in two junctions
simultaneously, i.e. the voltage grows across
both junctions. 

\begin{figure}
\centerline{\psfig{figure=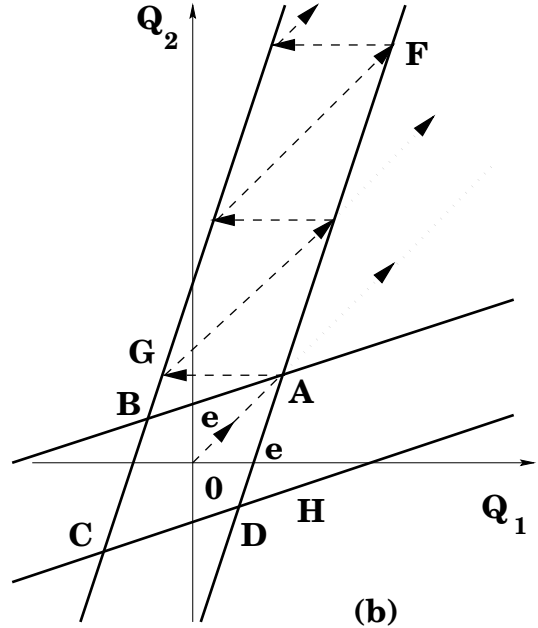,width=7.cm}}
\vspace{0.3cm}
\caption{The same as Fig.~2, $C_0 \ne 0$. The first Brillouin zone is
a romb ABCD. In the weak current 
regime the system moves along the dashed line BF.} 
\label{brill}
\end{figure}

The behavior is quite different in the presence of a finite
self-capacitance $C_0$. The system again starts from the
point $(0,0)$ and reaches the critical surface in the point A$=(e\xi,
e\xi)$, $\xi = (1-\alpha)^{-1}$, $1 < \xi \le 2$ (Fig.~3), {\em
beyond} the point $(e,e)$. Consequently, it has only three
possibilities of further motion. The first one (with probability
$(1-\lambda^2)/2$, where $\lambda$ is given by the same expression
(\ref{pro1}) with $E_C = e^2/2(C+C_0)$) is to be reflected to
the point G$=(e(\xi-2),e\xi)$. It corresponds to Zener tunneling in 
junction 2. The subsequent evolution of the system is voltage
growth across junction 2 and Bloch oscillations in junction 1. 
The equivalent possibility is to be reflected to the point H$(=e\xi,
e(\xi-2))$. Finally, the last allowed process (with probability
$\lambda^2$) is to continue the motion along the line $(It,It)$
without any reflection, i.e. Zener tunneling takes place in both
junctions. Note that the simultaneous occurrence of Bloch oscillations
in both junctions is impossible. This is a consequence of a lowered
symmetry: while the case $C_0=0$ has the same symmetry group as a
square, the introduction of $C_0$ lowers the latter to that of a romb.

Thus, the situation differs qualitatively from that of a single
junction. Still two scenarios exist, low and high
currents. The second one yields the mean voltage $2V_0$, which one
could expect. However, the low current scenario leads now to a finite
mean voltage $V_0$, in contrast to the single-junction
case. The time $t_s \sim 2e\Delta/IE_C \ll t_Z$, required for this
metastable value to establish, is very short. Hence at the beginning
of experiment the mean value $V_0$ is observed. The time of order
$t_Z$, which can be extremely long, is required to obtain the stable
value $2V_0$. 

\begin{figure}
\centerline{\psfig{figure=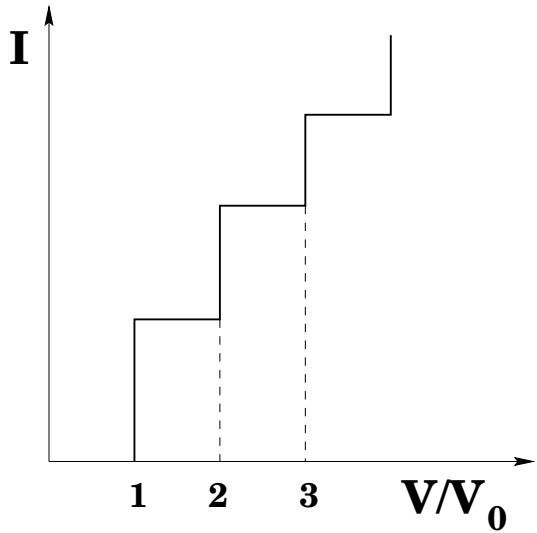,width=7.cm}}
\vspace{0.3cm}
\caption{A sketch of voltage-current characteristic of the system of
coupled junctions. All fine structure of oscillations is eliminated,
and only the mean value $\bar V$ in units of $V_0$ is
shown.}
\label{ivchar}
\end{figure}

The above approach can be generalized to the case of
the Josephson chain consisting of $N+1$ grains and $N$ junctions. The
relevant variables are phases of 
junctions $\theta_i = \phi_i - \phi_{i+1}$, $1 \le i \le N$, and the
center-of-mass coordinate $\theta_{N+1} = (N+1)^{-1}(\phi_1 + \phi_2 +
\dots + \phi_{N+1})$, with $\phi_i$ being the phases of the
grains. The problem is equivalent to the motion of an 
electron in an $N$-dimensional periodic potential; the chain is
represented by a point $\mbox{\bf Q}$ in the space of junction
charges $Q_i, i = 1, \dots, N$. A charge-neutral system moves along
the lines  
$$(Q_1 + 2eN_1, Q_2 + 2en_2, \dots, Q_N + 2en_N), n_k \in \cal Z.$$
It starts from the coordinate origin and can jump between the lines only
as a result of ``Bragg reflection'' on the critical surfaces, given by ($l
= 1, \dots, N$)
\begin{equation} \label{crit1}
\sum_i Q_i (C^{-1})_{il} = \pm e (C^{-1})_{ll},
\end{equation}
where the $N \times N$ capacitance matrix is given by
\begin{equation} \label{matrN}
C_{kl} = C_{lk} = C\delta_{kl} + C_0 \frac{k(N+1-l)}{N+1}; \ \ k < l. 
\end{equation}

An analysis for arbitrary relation between capacitances $C$ and
$C_0$ would require the inversion of the capacitance matrix $\hat C$. 
However, both cases $C=0$ and $C_0 \ll C$
allow an analytical solution and show the same symmetry lowering
causing a finite (metastable) voltage $V_0$ across the chain. 

Thus, the current-voltage characteristics can be described as
follows. The stable value of the mean voltage for {\em any} value of
current is $NV_0$. However, the time required to achieve this stable
voltage is astronomical (exponentially long) for low currents, and
may exceed the duration 
of the experiment $t_e$. Hence one
observes the metastable value $V_0$. The main effect of an increasing
current is a decrease of the time $t_Z$. For high currents one has 
$t_Z \ll t_e$ and observes the stable voltage. There are some
intermediate regimes between, and metastable voltages $kV_0$,
$k=2,3,\dots, N-1$ can be measured (Fig.~4). This dependence 
is essentially a low-voltage that observed in the experiments
\cite{Jap}.  

In conclusion, we predict that a current-biased chain of $N$
capacitively coupled Josephson junctions 
always is in a finite voltage $\bar V \geq 2\Delta /e$ state
no matter how low the external current is \cite{foot3}. 
This conclusion differs qualitatively from that reached for
a single Josephson junction. It is based on simple symmetry arguments.
While for vanishing on-site capacitance $C_0=0$ the structure of 
the Brillouin zone for a Josephson
``particle'' corresponds to that for an N-dimensional cubic crystal,
the crystal symmetry is lowered for any non-zero $C_0$. As a result
Bloch oscillations in the lowest Brillouin zone cease to occur and
the system ``jumps'' to higher bands due to Zener tunneling. This
value of the voltage is metastable, but can be measured experimentally
since the time scale required to observe the stable value $NV_0$ is
very long for low currents. We found a staircase-like shape of the I-V
curve of the system and argue that our predictions can explain the observed
low-voltage behavior of Josephson junction arrays \cite{Jap}.

The authors are grateful to R.~von Baltz, D.~Esteve, C.~Fuchs,
D.~Haviland, A.~Kanda, S.~Kobayashi, L.~S.~Kuzmin, and K.-H.~Wagenblast for
useful discussions. One of us (G.~S.) acknowledges the A.~v.~Humboldt
award of the Academy of Finland and the hospitality of the Helsinki
University of Technology during the progress of this work. Two of us
(Y.~B. and A.~Z.) acknowledge the support by the Alexander von
Humboldt Foundation. The project was supported by the German Science
Foundation within the SFB 195.


\begin{thebibliography}{99}
\bibitem{foot1} For a cosine potential the gaps between higher energy 
bands are narrower
(and hence the probability of Zener tunneling is higher) than
that between the two lowest ones.
\bibitem{foot2} Our considerations can be generalized  
to the configuration with gate voltages inducing non-zero
gate charges on superconducting grains. In this case fractional parts of 
these charges determine the initial position of the system 
in the first Brillouin zone. No new physical effects arise from this 
complication and all our main conclusions remain valid also for
the case $Q_3 \neq 0$.
\bibitem{foot3} In order to relate our simple model to the real
experimental situations, one should connect the system with two
additional junction to two reservoirs. It can be easily shown that 
the minimal system in order to observe the predicted effect is shown
on Fig.~1b.
\end{thebibliography}
\end{document}